# Generalized master curve procedure for elastomer friction taking into account dependencies on velocity, temperature and normal force

Valentin L. Popov[1], Lars Voll[1], Stephan Kusche[1], Qiang Li[1] and Svetlana V. Rozhkova[2]

[1]Berlin University of Technology, Berlin, 10623, Germany
[2]Tomsk Polytechnic University, Tomsk, 634050, Russia
v.popov@tu-berlin.de

**Abstract**
In the sliding contact of elastomer on a rigid substrate, the coefficient of friction may depend on a large number of system and loading parameters, including normal force, sliding velocity, shape of contacting bodies, surface roughness and so on. It was argued earlier that the contact configuration is determined more immediately through the indentation depth than the normal force, and thus the indentation depth can be considered as one of "robust governing parameters" of friction. Both models of friction of simple shapes and fractal surfaces demonstrate that the coefficient of friction of elastomers should be generally a function of dimensionless combinations of sliding velocity, surface gradient, relaxation time and size of micro-contacts. The relaxation time does depend only on temperature and the surface slope and the size of micro contacts mostly on the indentation depth. Based on this general structure of the law of friction, we propose a generalized master curve procedure for elastomer friction where the significant governing parameter – indentation depth (or normal force) was taken into account. Unlike the generation of the classical master curve by horizontal shifting of dependence "friction - logarithm of velocity" for different temperatures, in the case of various indentation depth the shifting in both horizontal and vertical direction is required. We experimentally investigated coefficient of friction of elastomer on sliding velocity for different indentation depths and temperatures, and generated a 'master curve' according to this hypothesis.

**Introduction**

Friction exists everywhere in both nature and man-made objects. It can be beneficial, such as book flipping or product transportation on conveyor system [1][2]; sometimes friction is expected to be reduced because it causes the unwanted loss of energy [4]-[6]. A simple but today still widely used law of friction is Amontons' law which states that friction force is proportional to the normal force, the ratio is known as coefficient of friction [7]. However, since Coulomb it has been already known that coefficient of friction may depend also on material, normal load, system size, time, sliding velocity and so on [8]-[10]. These factors will have a significant influence on friction of elements in precision instruments. Especially in micro/nano- scale system, due to the high surface area-to-volume ratio and surface interaction, the frictional behavior is more complicated and usually cannot be described by a general law [11]-[13].

It is well known that for rubber like viscoelastic materials friction in a contact with a hard counter surface is velocity-dependent [14]. A large number of studies have shown that friction between elastomer and rigid surface depends on all loading parameters and material parameters, in particular strongly on sliding velocity, normal load and temperature [15]-[16].. In 1963 Grosch firstly presented the so-called master curve procedure [20] which since then became a standard procedure in rubber industry. It is based on the shifts of dependencies of the coefficient of friction on logarithm of velocity measured at different temperatures in horizontal direction by the WLF (Williams, Landel, Ferry) factors [21], to a single continuous 'master curve' which describes the frictional behavior of rubber in a large velocity range and at the same time at different temperatures. This method has been widely applied in analysis of elastomer friction [22][23]. Similar master curve for stress (or shear modules)–frequency dependence investigated by dynamic mechanical analysis and dielectric spectroscopy is also a common method to study and characterize the viscoelastic properties of polymer material, including determination of complex modulus and glass transition temperature [24][25]. Usually these master curves of friction-velocity and stress-frequency dependences are generated by horizontal shifting because only one further parameter, temperature is considered. However it is found that a vertical shift procedure has to be introduced into the generation of master curve in some cases due to the complex interaction of internal structures [26]. Furthermore, influence of other loading parameters such as normal force, have been rarely taken into account [27], in spite of the obvious practical importance.

Recently it was argued that, although there are a large number of parameters which affect the friction, it occurs to be possible to reduce the number of parameters by choosing corresponding parameter combinations which mostly directly and robustly determine friction. It was argued that one of such parameters is the indentation depth [28]. A simple example of this robustness is well known: the relation between contact radius and indentation depth in a non-moving sphere contact, $a = \sqrt{Rd}$ is determined solely by the indentation depth (for a given shape) independently of the elastic properties of the medium. In addition, for frictional contact it is already theoretically and experimentally shown that the maximum pre-sliding displacement is function of only coefficient of friction and indentation depth, $u_{x,\max} = 1.5\mu d$ [29][30]. Based on discussion of these robust governing parameters in [27][28], we suggest a generalized master curve procedure of elastomer friction taking into account sliding velocity, temperature and now also the normal force. These loading parameters can be easily experimentally realized and controlled. For verification we carried out corresponding experiments using a tribometer with controlled indentation depth.

**Results**

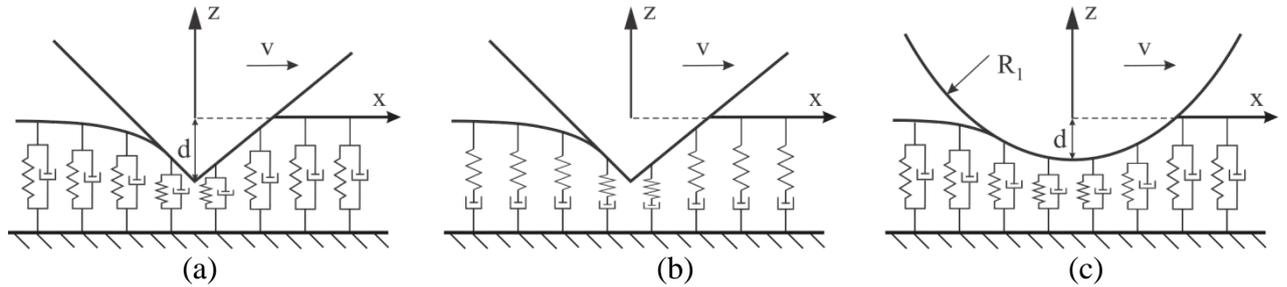

**Fig. 1** Contacts between rigid indenters and viscoelastic foundations.

Just for illustrating the main idea of the new suggested method let us consider several simple examples of friction of a rigid indenter against an elastomer. As examples, we consider two contacts of conical indenters with viscoelastic media having different rheology and one example of a parabolic indenter in contact with a Kelvin body. For the case of simplicity an illustration we consider one-dimensional contacts with elastic foundations. However, the results have been confirmed through complete three dimensional boundary element simulations [31]. The rigid indenter is pressed against a viscoelastic foundation with indentation depth $d$ and then slides tangentially with a constant velocity $v$. If the indenter has a conical profile $z = g(x) = c|x|$, and viscoelastic material is modelled as a combination of series of Kelvin-element, as shown in Fig. 1a, the coefficient of friction between these the contacting bodies is given by the equation [32]

$$\mu = \nabla z \frac{\left[2\left(\frac{\nabla z \cdot v\tau}{d}\right) - \frac{1}{2}\left(\frac{\nabla z \cdot v\tau}{d}\right)^2\right]}{\left[1 + \frac{1}{2}\left(\frac{\nabla z \cdot v\tau}{d}\right)^2\right]}, \quad (1)$$

where $\nabla z$ is surface gradient and equal to profile coefficient $\nabla z = c$, $\tau$ is relaxation time of the elastomer. If Kelvin-elements are replaced by Maxwell-elements (Fig. 1b), then the result is given by [32]

$$\mu = \nabla z \cdot \frac{\dfrac{d}{\nabla z \cdot v\tau} - 2\left(1 - e^{-\frac{d}{\nabla z \cdot v\tau}}\right) + \ln\left(2 - e^{-\frac{d}{\nabla z \cdot v\tau}}\right)}{\dfrac{d}{\nabla z \cdot v\tau} - \ln\left(2 - e^{-\frac{d}{\nabla z \cdot v\tau}}\right)}. \tag{2}$$

For sliding contact of spherical indenter (radius $R_1$) whose profile is described as $g(x) = x^2/R_1$ with a Kelvin-element foundation (Fig. 1c), the coefficient of friction is given by [32]

$$\mu = \nabla z \frac{\xi\left[2 - 3\xi - 2\xi^3 + 2(1+\xi^2)^{3/2}\right]}{\left[1 - \xi^3 + (1+\xi^2)^{3/2}\right]^{4/3}}, \tag{3}$$

where $\xi = \dfrac{v\tau}{(2R_1 d)^{1/2}} \approx \dfrac{v\tau}{a}$ and $a$ is contact radius.

In all considered cases, the dependence of the coefficient of friction on parameters has the same structure:

$$\mu = \nabla z \cdot \Psi\left(\frac{v\tau}{a}\right). \tag{4}$$

Similar results have been obtained also for rough fractal contacts [28]. This simple equation is of course just representation of the old and well known results that the rheological contribution to elastomer friction is a product of the rms slope of the surface roughness and the theological factor depending on the product of the characteristic frequency $v/a$ and the relaxation time of the elastomer [17]. The geometrical parameters of the contact configuration, $\nabla z$ and the characteristic size of the "asperities", $a$, can of course depend on the indentation depth. But according to the Archard's idea, the geometrical parameters of contact asperities are only weakly dependent of loading conditions. As no loading parameter is involved in (4), this equation will also be valid for multi-contact configuration.

In Eq. (4), the parameters surface gradient $\nabla z$ and size of local contact $a$ depend not only on the indentation depth but also on the surface roughness. It is practically impossible to measure these properties. But for our purpose this is not needed, only the general structure given by Eq. (4) is important. It is sufficiently to know that these quantities are determined solely by the indentation depth [33]-[35]. The relaxation time $\tau$ describes the rheology of the material and is dependent of temperature. Therefore, Eq. (4) can be rewritten as function of indentation depth and temperature

$$\mu = \nabla z(d) \cdot \Psi\left(\frac{v \cdot \tau(T)}{a(d)}\right). \tag{5}$$

In a double logarithmic scale it takes the form

$$\log \mu - \log \nabla z(d) = \Phi\left(\log v + \log \tau(T) - \log a(d)\right), \tag{6}$$

where $\Phi$ is a new function $\Phi(\cdot) = \log \Psi(\exp(\cdot))$. This form suggests a generalized master curve procedure for determining the coefficient of friction as a function of sliding velocity, temperature and the indentation depth. According to (6), the curves of $\log \mu$ vs $\log v$ for different temperatures and indentation depths will have the same form only shifted along the vertical or horizontal axis. If we shift the curves horizontally by a factor $\log \tau(T) - \log a(d)$ and vertically by a factor $\log \nabla z(d)$, the should form one single "master curve".

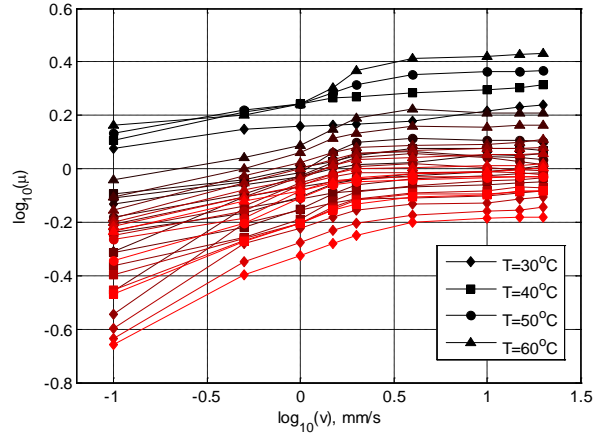

**Fig. 2** Measured coefficient of friction in the experiment. The colors changing gradually from dark to red indicate the 8 linearly increasing indentation depths from 0.1 to 0.8 mm.

For validation of the proposed procedure (6) we experimentally investigated the coefficient of friction of a steel sphere sliding on an elastomer plate for different sliding velocities, indentation depths and temperatures. The following ranges of three parameters were investigated: 8 indentation depths from 0.1 to 0.8 mm, 9 sliding velocities from 0.1 to 20 mm/s and 4 operation temperatures from 30 to 60 °C. The results of experiment measurements are shown in Fig. 2, where double logarithmic coordinates are used, and the 8 increasing indentation depths are indicated by colors changing gradually from dark to red. It can be seen that the coefficient of friction increases generally with the sliding velocity then reaches a plateau. For a larger indentation depth, the friction becomes smaller, and also the higher the temperature is, the smaller the coefficient of friction becomes.

Now we carry out the 'shifting'-procedure. According to (6) the shift of curves in Fig. 2 will generate a 'master curve', which can be described as logarithmic dependence of $\bar{\mu}$ on $\bar{v}$:

$$\log_{10} \bar{\mu} = \log_{10} \mu - A(d_i), \qquad (7)$$

$$\log_{10} \bar{v} = \log_{10} v_n + B(d_i) + C(T_j), \qquad (8)$$

where the shift factors $A$ and $B$ are functions of indentation depth, $A = f(d_i)$, $B = f(d_i)$, with $i = 1\ldots 8$, $C$ is function of temperature, $C = f(T_j)$ with $j = 1\ldots 4$, $n$ indicates 9 sliding velocities.

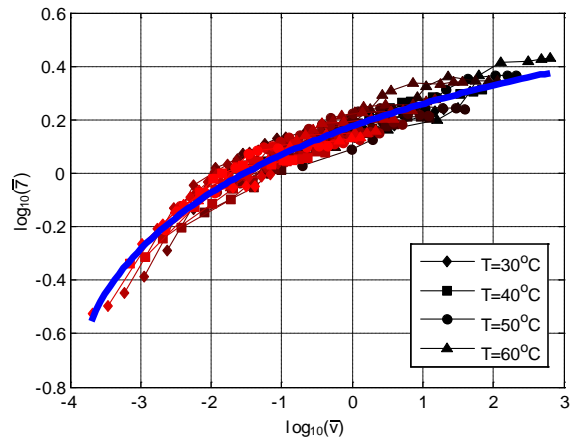

**Fig. 3** 'Mastercurve' and fitting line. . The curve for d=0.1 and T=30°C is fixed.

To obtain a 'best master curve', we used a fitting equation of the form

$$M + M_0 = \frac{\xi_0 \cdot (V + V_0)^{\xi_1}}{1 + \xi_2 \cdot (V + V_0)^{\xi_1}}, \qquad (9)$$

where $M$ is a function of expected form 'increasing with a following plateau' shape, and its value is expected to be close to $\log_{10} \bar{\mu}$, $V = \log_{10} \bar{v}$, $\xi_0$, $\xi_1, \xi_2$ $V_0$, and $M_0$ are unknown constants. The appropriate master curve will be evaluated by the calculation of the minimal error as following

$$\delta = \sum \left(\log_{10} \bar{\mu} - M\right)^2. \qquad (10)$$

The error $\delta$ is the function of the following set of variables: $\{d_i, T_j, v_n, A(d_i), B(d_i), C(T_j), \xi_0, \xi_1, \xi_2, V, M_0\}$. The search for the minimum of this function was carried out using methods of mathematical statistics. The generated master curve is shown in Fig. 3 with error $\delta_{\min} = 0.4556$, and the following values of the fitting parameters: $M_0 = 0.6264$, $\xi_0 = 0.5017$, $\xi_1 = 0.7591$ and $\xi_2 = 0.2615$. As reference, the values $d = 0.1$ mm and $T = 30°$ were chosen. The blue curve is the corresponding fitting line according to Eq.(10). The obtained dependences of shift factors $A$, $B$ on indentation depth, and factor $C$ on temperature are presented in Fig. 4.

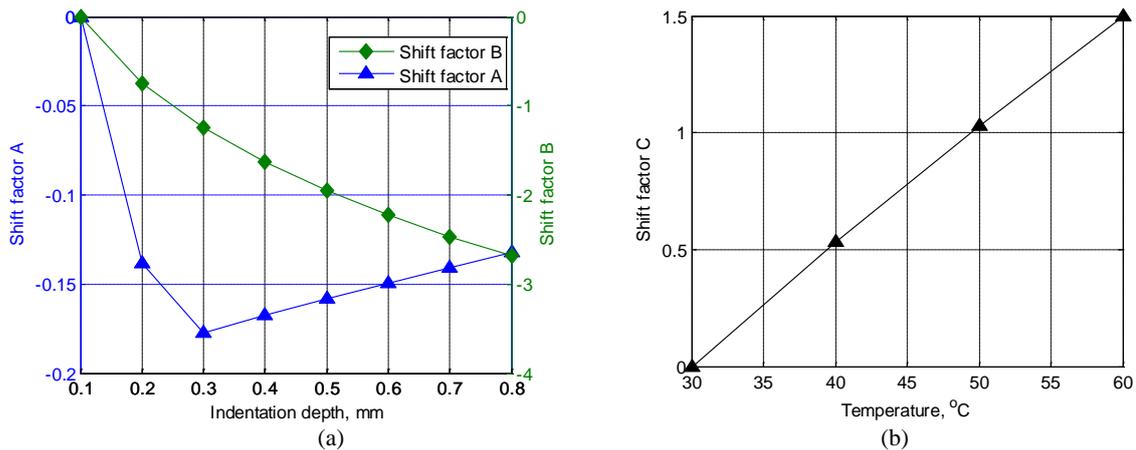

**Fig. 4** Dependence of shift factor A, B on indentation depth and C on temperature.

**Discussion**

We analysed the coefficient of friction of elastomers contacting with different shaped indenters. The solutions shown that the coefficient of friction depends on only two dimensionless parameters: surface gradient and a combination of sliding velocity, relaxation time and local contact size. This result represents a general form valid also for rough contacts. Based on that we suggested a master curve procedure where the indentation depth was considered as the most robust parameter characterizing normal loading. In the experiment we measured the dynamic friction between a rubber band and a steel ball. The obtained dependence of coefficient of friction on sliding velocity for different temperatures and indentation depths generated a master curve by shifting in both horizontal and vertical directions.

## Methods

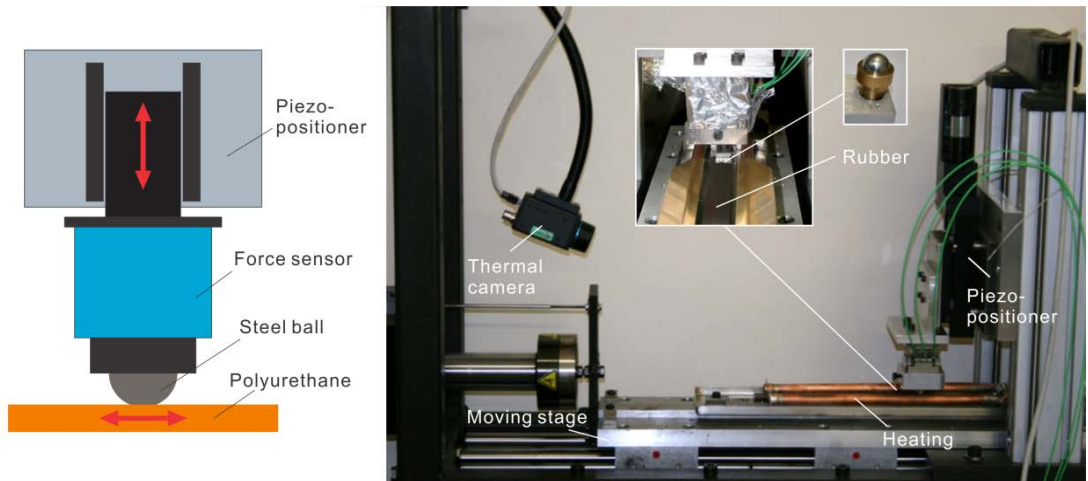

**Fig. 5** Experimental set-up for measurement of sliding friction between a steel ball and a rubber band.

**Experimental investigation.** The experimental set-up for measurement of elastomer friction is shown in Fig. 5. The rubber band with a size of $300 \times 50 \times 8$ mm was fixedly glued to a moving stage which is controlled by a hydraulic actuator. The material of the band is polyurethane and its shear modulus is about $3$ MPa and yield modulus about $48$ MPa. The steel ball has a radius $R = 6$ mm and is much harder than the rubber band. It was mounted on a 3D force sensor and they are together controlled by a piezo-positioner to move vertically. The operation temperature was controlled by heating equipment. The heating tubes were installed around the rubber band to keep a stable temperature. The temperature changes of the rubber surface in experiment were controlled by a thermal camera.

During the measurement the steel ball was pressed into the rubber band with a given indentation depth by piezo-positioner, and then the rubber band was moved horizontally with a constant velocity. The frictional force and normal force were measured by a 3D force sensor and the coefficient of friction was therefore easily calculated. Measurements were carried out in the velocity range from 0.1 to 20 mm/s for indentation depth from 0.1 to 0.8 mm and temperature from 30 to 60 °C. Data for any parameter set (velocity, indentation depth and temperature) were averaged over six measurements.

## Author contributions statement

V.L.P. and S.V.R. conceived of the study. V.L.P. coordinated the study and carried out the theoretical analysis. L.V. designed the experimental setup, carried out the measurement and collected the data. S.K. and Q.L. carried out the statistical analyses. V.L.P. and Q.L. drafted the manuscript. All authors reviewed the manuscript.

## Additional information

Competing financial interests: The author declares no competing financial interests.